\documentclass{PoS}

\title{Studies of exotic hadrons at Tevatron and LHC}

\ShortTitle{Studies of exotic hadrons at Tevatron and LHC}

\author{\speaker{Vincenzo Chiochia}\\
        Physik-Institut, Winterthurerstr. 190, 8057 Zurich, Switzerland\\
        E-mail: \email{Vincenzo.Chiochia@cern.ch}}

\abstract{In these proceedings we summarise the recent studies of exotic hadrons performed at the LHC and Tevatron. The X(3872) production cross section and mass measurements are reported, along with the determination of its quantum numbers, the searches for decays to $\mathrm{p\bar{p}}$, and for partner states in the bottomonium sector. Furthermore, we discuss the Y(4140) and Y(4274) structures, observed by several experiments in the $J/\psi\phi$ invariant mass spectrum. }

\FullConference{XV International Conference on Hadron Spectroscopy-Hadron 2013\\
		4-8 November 2013\\
		Nara, Japan }

\begin{document}

%

\section{Studies of the X(3872) state and searches for the bottomonium partners\label{sec:X3872xsec}}

The first observation of the X(3872) state was reported by the Belle collaboration in 2003 using $\mathrm{B^+}\rightarrow J/\psi\mathrm{K^+}\pi^+\pi^-$ decays~\cite{Choi:2003ue} and later confirmed by CDF, D0 and BaBar~\cite{Acosta:2003zx,Abazov:2004kp,Aubert:2004ns}. Although considerable experimental efforts have been dedicated to the study of this state, more than ten years after the discovery its nature  is still uncertain: the X(3872) could be a conventional charmonium state~\cite{Skwarnicki:2003wn}, or an unconventional state such as a loosely bound $\mathrm{D}^{*0}\mathrm{\bar{D}}^0$ molecule~\cite{Tornqvist:2004qy},  a tetraquark state~\cite{Maiani:2004vq}, or a charmonium-molecule admixture~\cite{Hanhart:2011jz}. The state is abundantly produced at hadron colliders, thus cross section measurements can help constraining the particle nature and provide a test of the QCD approaches to X(3872) production. 

\begin{figure}[htbp]
\begin{center}
\includegraphics[width=6.5cm]{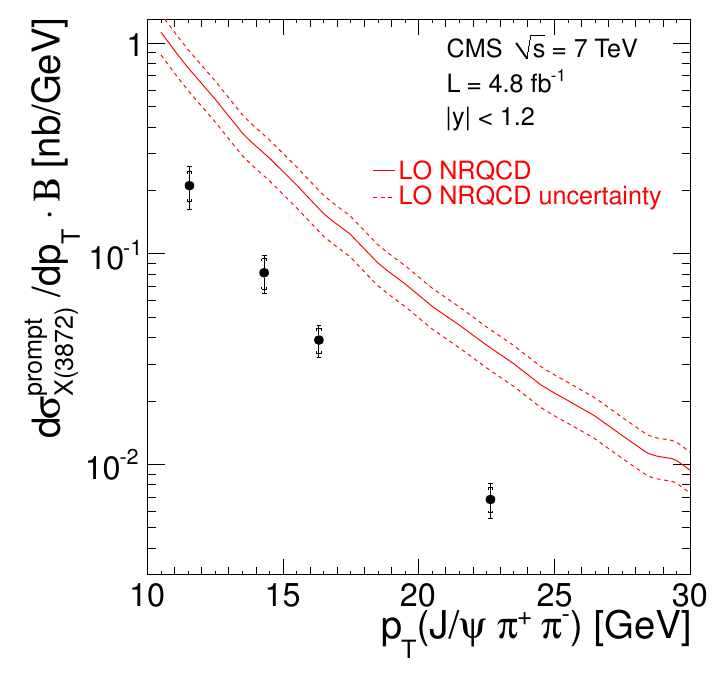}
\includegraphics[width=8.5cm]{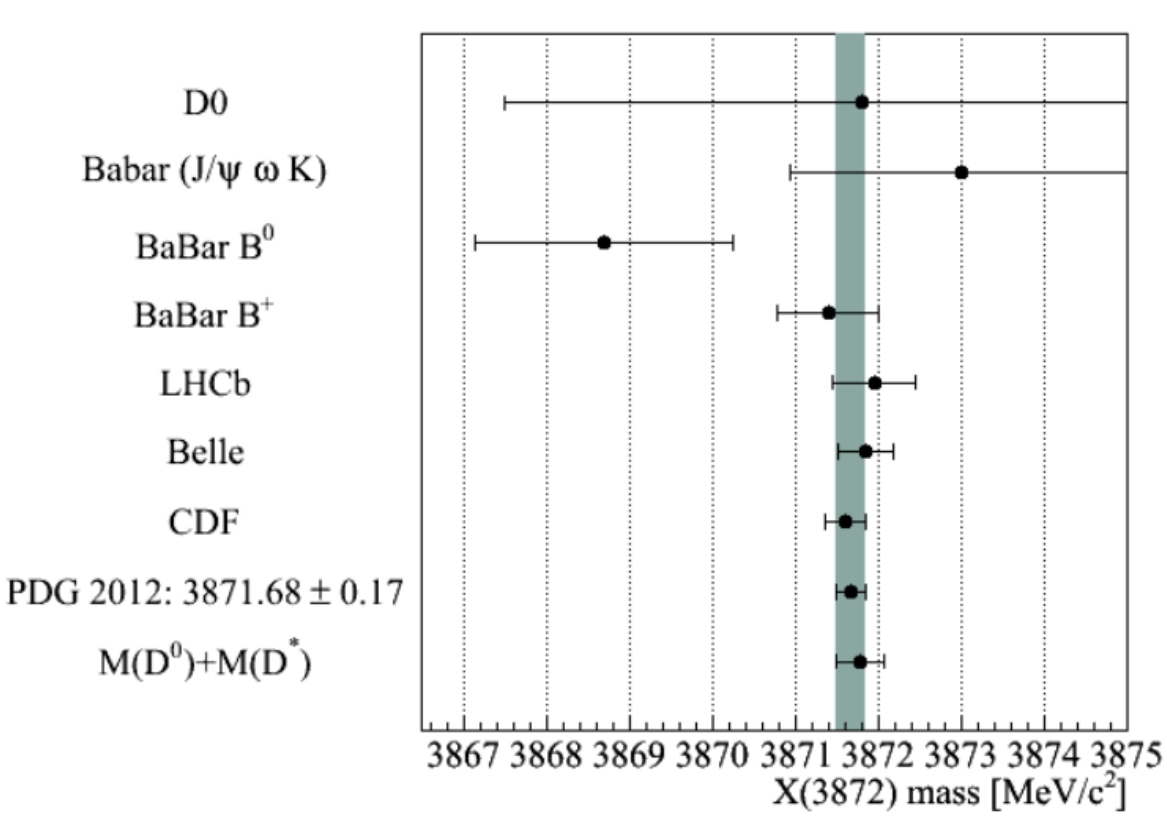}
\caption{{\it Left:} Cross section times branching fraction for X(3872) prompt production as function of $p_T$ as measured by CMS, compared to NRQCD predictions. {\it Right}: Summary of the X(3872) mass measurements.}
\label{fig:CMSX3872}
\end{center}
\end{figure}

Production cross section measurements in $\mathrm{pp}$ collisions at $\sqrt{s}=7$~TeV were recently conducted at CMS and LHCb in the central and forward rapidity regions, respectively~\cite{Chatrchyan:2013cld,Aaij:2011sn}. 
The CMS measurement, based on an integrated luminosity of 4.8~fb$^{-1}$, was performed in a kinematic range corresponding to X(3872) candidates with transverse momentum $10 < p_T < 50$~GeV and rapidity $|y| < 1.2$. X(3872) candidates were reconstructed in $J/\psi(\mu^+\mu^-)\pi^+\pi^-$ decays. The total yield extracted from a fit to the invariant mass distribution was about 12\,000 candidates. The prompt X(3872) cross section was determined from the ratio to the prompt $\psi(2S)$ cross section and fraction of X(3872) from b decays, $f_{X(3872)}^B$. The latter was extracted from the decay lifetime distribution and found to be significantly smaller than that for the $\psi\mathrm{(2S)}$, $f_{X(3872)}^B=0.263 \pm 0.023 \pm 0.016$, with no significant dependence on $p_T$. The candidate yields were measured in bins of $p_T$ and corrected for detector efficiency and luminosity under the assumption that the quantum numbers are $J^{PC}=1^{++}$. The resulting prompt differential cross section as a function of $p_T$ is shown in Fig.~\ref{fig:CMSX3872} ({\it left}). Measurements were compared to non-relativistic QCD (NRQCD) predictions, which were found to be about four times above the data. CMS also found that a large fraction of the total X(3872) cross  section, about 75\%, is by given by prompt production and it is roughly independent on $p_T$. Furthermore, the dipion invariant mass spectrum was found to be consistent with  the intermediate $J/\psi\rho(\pi^+\pi^-)$ decay hypothesis.

Forward X(3872) production was measured by LHCb, in the kinematic range $5 < p_T < 20$~GeV and rapidity $2.5 < |y| < 4.5$ with $J/\psi(\mu^+\mu^-) \pi^+\pi^-$ decays. Using an integrated luminosity of 34.7~pb$^{-1}$ a fit to the invariant mass distribution yielded $565\pm62$ candidates. The production cross section multiplied by the branching fraction was found to be $5.4 \pm 1.3 \mathrm{(stat)} \pm 0.8 \mathrm{(syst)}$~nb, which is also in this case considerably below the NRQCD prediction of $13.0\pm2.7$~nb. LHCb also measured the X(3872) mass from an extended unbinned maximum likelihood fit of the reconstructed $J/\psi\pi^+\pi^-$ mass in the interval $3.60 < M < 3.95$~GeV/c$^2$. The signal was described with a non-relativistic Breit-Wigner (BW) function convolved with a Gaussian resolution function~\cite{Aaij:2011sn}. The X(3872) width was fixed to zero in the fit and the ratio of the mass resolutions for the X(3872) and the $\psi\mathrm{(2S)}$ was set to the value estimated from the simulation. The measured mass is $3871.95 \pm 0.48\mathrm{(stat.)}\pm 0.12\mathrm{(syst.)}$~MeV/c$^2$, where the systematic uncertainties are largely associated to the momentum scale. Given the small size of the analyzed data sample, LHCb has still good prospects for reducing the statistical uncertainty by using the complete dataset. A summary of the available X(3872) mass measurements is shown in Fig.~\ref{fig:CMSX3872} ({\it right}), where the current world average is $3871.68 \pm 0.17$~MeV/c$^2$ and the most precise determination is given by CDF.

For the $\mathrm{D}^{*0}\mathrm{\bar{D}}^0$ 'molecule' interpretation to be valid, the X(3872) mass is expected to be below the sum of the $\mathrm{D}^{*0}$ and $\mathrm{\bar{D}}^{0}$ masses. Available data are inconclusive, indicating that such a bound state would have a binding energy of $0.16 \pm 0.32$~MeV/c$^2$. To clarify the situation LHCb has also performed a precise measurement of the $\mathrm{D}^{0}$ mass using $\mathrm{D}^0 \to \mathrm{K}^+\mathrm{K}^-\mathrm{K}^-\pi^+$ decays~\cite{Aaij:2013uaa}. This decay mode is characterized by a low Q-value and the small momentum available for the final state particles helps reducing the systematic uncertainties associated to the momentum scale of the detector. The mass was extracted from an extended unbinned maximum likelihood fit to the invariant mass distributions, using an integrated luminosity of 1~fb$^{-1}$. The value measured from about 850 signal candidates is $1864.75 \pm 0.15\mathrm{(stat)}\pm 0.11\mathrm{(syst)}$~MeV/c$^2$, which has a similar precision of the available CLEO result. The corresponding value of the  binding energy for an hypothetical $\mathrm{D}^{*0}\mathrm{\bar{D}}^0$ dimeson is $0.09 \pm 0.28$~MeV/c$^2$, supporting the fact that such a 'molecule' would be an extremely loose bound state.

Early CDF studies constrained the X(3872) quantum numbers to the $J^{PC}=1^{++}$ or $2^{-+}$ options~\cite{Abulencia:2006ma}. These hypotheses were later confirmed by Belle and BaBar but the size of the analyzed datasets could not establish one of the two options with a sufficient degree of statistical confidence~\cite{Choi:2011fc,delAmoSanchez:2010jr}. LHCb recently reported the first five-dimensional angular analysis of the $\mathrm{B}^+ \to \mathrm{X(3872)K}^+$ decay with $\mathrm{X(3872)} \to J/\psi(\mu^+\mu^-) \pi^+\pi^-$, using an integrated luminosity of 1~fb$^{-1}$~\cite{Aaij:2013zoa}. A fit to the invariant mass distribution yielded $313\pm 26$ $\mathrm{B}^+$ candidates and the quantum numbers were extracted from the angular correlations of the final state particles. The $1^{++}$ and $2^{-+}$ hypotheses were discriminated through a likelihood ratio test, where the probability density function for each hypothesis were defined in a 5-dimensional angular space and assuming the dipion system to originate from an intermediate $\rho(770) \to \pi^+\pi^-$ decay. The measured value of the likelihood ratio corresponds to a $8.2\sigma$ rejection of the $2^{-+}$ hypothesis. This result rules out the interpretation of the X(3872) as conventional $\eta_{c2}$ state. Among the remaining possibilities are the $\chi_{c1}(2^3 P_1)$, which is disfavored by the X(3872) mass, a $\mathrm{D}^{*0}\mathrm{\bar{D}}^0$ molecule, a tetraquark state, or a charmonium-molecule mixture.

Further inside on the X(3872) nature can be obtained by studying its partial width to $\mathrm{p\bar{p}}$ decays. Searches for $\mathrm{X(3872)}\to \mathrm{p\bar{p}}$ decays are also crucial to predict the production rate at future $\mathrm{p\bar{p}}$ colliders~\cite{Lange:2010dt}. A search for $\mathrm{X(3872)}\to \mathrm{p\bar{p}}$ was performed by LHCb with $\mathrm{B^+} \to \mathrm{p\bar{p}} \mathrm{K^+}$ decays, using a data sample corresponding to an integrated luminosity of 1~fb$^{-1}$~\cite{Aaij:2013rha}. Particle identification requirements based on the RICH detector, were applied to the three-track final state and a signal yield of about 6950 candidates was determined from an unbinned extended maximum likelihood fit to the invariant mass distribution of the selected candidates. The signal yields for the charmonium contributions, $\mathrm{B^+}\to (c\bar{c})\mathrm{K^+}\to \mathrm{p\bar{p}}\mathrm{K^+}$, were determined by fitting the $\mathrm{p\bar{p}}$ invariant mass distribution of signal candidates within the $\mathrm{B^+}$ mass signal window. An unbinned extended maximum likelihood fit to the $\mathrm{p\bar{p}}$ invariant mass distribution was performed over the mass range 2400-4500~MeV/$c^2$. No evidence of the X(3872) signal was found and a 95\% CL upper limit on the ratio $\mathrm{Br}(\mathrm{X(3872)}\to \mathrm{p\bar{p}})/\mathrm{Br}(\mathrm{X(3872)}\to J/\psi\pi^+\pi^-)<2\times 10^{-3}$ was determined. This limit challenges some of the predictions associated to the molecular scenario and is close to the range of predictions for a conventional $\chi_{c1}$ state~\cite{Aaij:2013rha}.

In analogy to the X(3872) state, a bottomonium counterpart, named $\mathrm{X_b}$, would be expected to be a narrow state decaying to $\mathrm{\Upsilon(1S)}\pi^+\pi^-$. Its mass could be either close to the $\mathrm{B\bar{B}}$ or $\mathrm{B\bar{B}^*}$ thresholds or, more generally, in the range 10-11~GeV/c$^2$. The CMS collaboration recently reported the results of a search for a peak in the $\mathrm{\Upsilon\mathrm{(1S)}(\mu^+\mu^-)}\pi^+\pi^-$ invariant mass spectrum other than the known $\Upsilon\mathrm{(2S)}$ and $\Upsilon\mathrm{(3S)}$ resonances. The analysis is based on an integrated luminosity of 20.7~fb$^{-1}$ collected at $\sqrt{s}=8$~TeV~\cite{Chatrchyan:2013mea}. The mass spectrum was modelled with a Gaussian function and fits to the data were performed by shifting the mass of the hypothetical state by 10~MeV/c$^2$ and letting the signal strength float. The $\mathrm{X_b}$ intrinsic width was assumed to be smaller than the detector resolution and the $\mathrm{\Upsilon\mathrm{(2S)}\to \Upsilon\mathrm{(1S)}(\mu^+\mu^-)}\pi^+\pi^-$ decay was selected as normalisation channel to reduce the systematic uncertainties. The expected signal significance for the ratio $R=\sigma(\mathrm{pp \to X_b \to \Upsilon(1S)\pi\pi})/\sigma(\mathrm{pp \to \Upsilon(2S) \to \Upsilon(1S)\pi\pi})=6.56\%$, motivated by the X(3872) case, was found to be above $5\sigma$ over the full measured mass range. No significant excess was observed between 10 and 11 GeV/c$^2$ and the 95\% CL upper limit on $R$ was set in the range 0.9-5.4\% depending on the $\mathrm{X_b}$ mass.

%
%
\section{The Y(4140) and Y(4274) structures in the $J/\psi\phi$ system}

In 2009 the CDF collaboration reported the evidence for a narrow peak near threshold in the $J/\psi \phi\to \mu^+\mu^-\mathrm{K^+K^-}$ system by studying $\mathrm{B^+} \to J/\psi\phi \mathrm{K^+}$ decays, which was called Y(4140)~\cite{Aaltonen:2009tz}. The search was later updated by processing a larger dataset, corresponding to an integrated luminosity of 6~fb$^{-1}$, yielding about 115 $\mathrm{B^+}$ candidate decays~\cite{Aaltonen:2011at}. In the new analysis a fit to the $J/\psi\phi$ invariant mass distribution yielded $19\pm7$ candidate events at a mass of $4143.4\pm 2.9\mathrm{(stat)}\pm 0.6\mathrm{(syst)}$~MeV/c$^2$, corresponding to a significance above $5\sigma$. A second structure, the Y(4274), with a mass of $4274.4\pm 8.4\mathrm{(stat)} \pm 1.9\mathrm{(syst)}$~MeV/c$^2$ was also observed, with significance above $3\sigma$. 

The Y(4140) structure could be interpreted as the signature of a new exotic meson with a narrow width, decaying to a pair of quarkonium states ($c\bar{c}$ and $s\bar{s}$). Potential interpretations include a $\mathrm{D_s^{*+}D_s^{*-}}$ molecule, an exotic charmonium hybrid with $J^{PC}=1^{-+}$, a $c\bar{c}s\bar{s}$ tetraquark state, or a less exotic consequence of the $J/\psi\phi$ channel threshold. The narrow structures inspired searches in other experiments, such as the one performed by Belle in the two-photon process $\gamma\gamma \to J/\psi\phi$, that found no significant structure near threshold but observed a narrow structure, named X(4350), at $4350.6^{+4.6}_{-5.1}\mathrm{(stat)}\pm 0.7\mathrm{(syst)}$~MeV/c$^2$~\cite{Shen:2009vs}. According to Belle, the upper limits on the product of the two-photon decay width and branching fraction of $\mathrm{Y(4140)} \to J/\psi\phi$ disfavor the scenario in which the Y(4140) is a $\mathrm{D_s^{*+}D_s^{*-}}$ molecule with $J^{PC}=0^{++}$ or $2^{++}$.

The LHCb collaboration~\cite{Aaij:2012pz} searched for the Y(4140) structure in $\mathrm{B^+} \to J/\psi\phi \mathrm{K^+}$ decays using 370~pb$^{-1}$ collected at $\sqrt{s}=7$~TeV. The dataset yielded about 380 signal candidates and events within $\pm 15$~MeV of the $\phi$ mass were selected to search for the Y(4140) state. LHCb observed no narrow structure near the threshold in the mass difference distribution $M(J/\psi\phi)-M (J/\psi)$. The CDF fit model was used to quantify the compatibility of the two measurements, fixing the mass and width of the Y(4140) peak to the central values obtained by the CDF collaboration. Including the second structure in fit did not affect the Y(4140) yield significantly. The LHCb collaboration quantified the disagreement between the two experiments at $2.4\sigma$ level and set an upper limit on $\mathrm{Br(B^+ \to Y(4140)K^+)} \times \mathrm{Br(Y(4140) \to J/\psi\phi)}/ \mathrm{Br(B^+ \to J/\psi\phi K^+)}$ of 0.07 at 90\% CL. The corresponding upper limit at 90\% CL for the second structure was found to be 0.08.
\begin{figure}[htbp]
\begin{center}
\includegraphics[width=6.7cm]{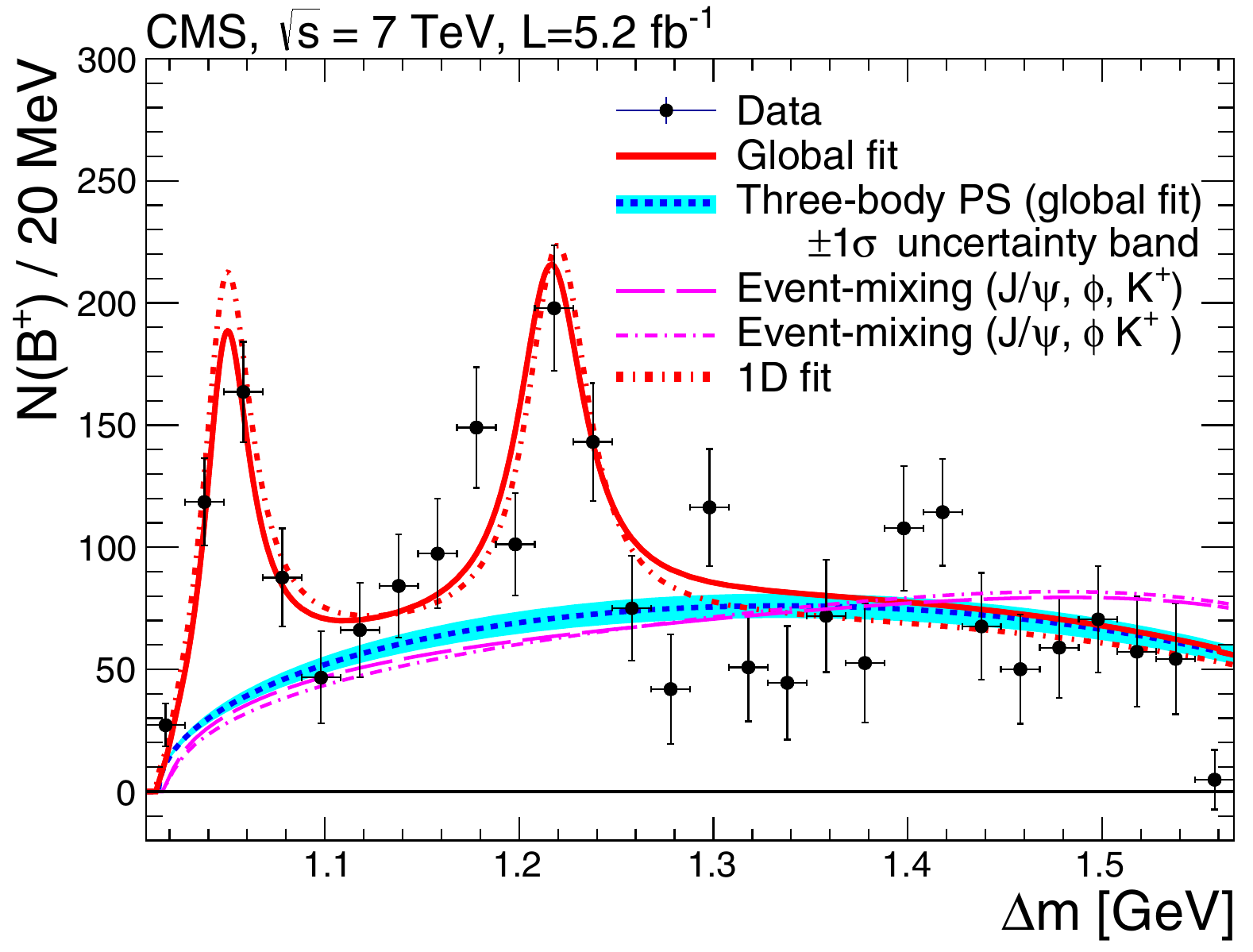}
\includegraphics[width=7.3cm]{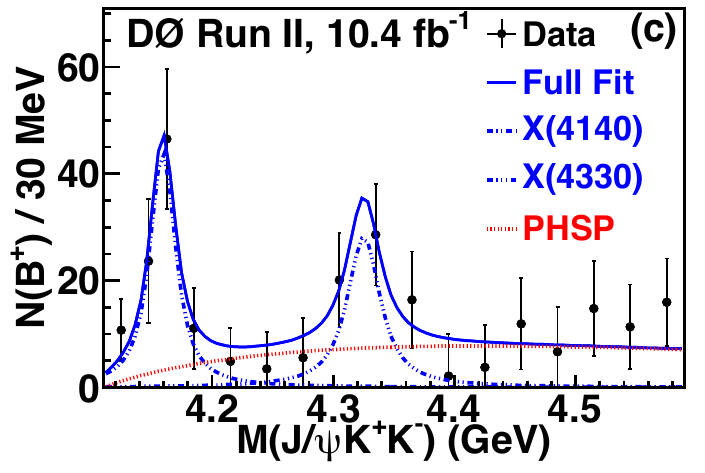}
\caption{$\mathrm{B^+}$ yields as function $\Delta M$ and $J/\psi\mathrm{KK}$ mass as measured by CMS ({\it left}) and D0 ({\it right}). Fits to the data assuming two narrow structures and a phase space background are also shown.}
\label{fig:CMS-D0-Y4140}
\end{center}
\end{figure}

The most recent searches  were reported by the CMS and D0 collaborations~\cite{Chatrchyan:2013dma,Abazov:2013xda}. CMS analyzed a data sample corresponding to 5.2~fb$^{-1}$ which yielded the largest number of $\mathrm{B^+}$ candidates to date, about 2\,480. To investigate the $J/\psi\phi$ invariant-mass distribution, the $J/\psi\phi\mathrm{K^+}$ candidates were divided into 20 MeV-wide $\Delta m = m(\mu\mu\mathrm{KK})-m(\mu\mu) $ intervals, and the $J/\psi\phi\mathrm{K^+}$ mass distributions for each interval were fit to extract the $\mathrm{B^+}$ signal yield in that interval, as shown in Fig.~\ref{fig:CMS-D0-Y4140} ({\it left}). CMS observed two peaking structures above the simulated three-body phase-space distribution. A first structure with mass $4148.0\pm 2.4\mathrm{(stat.)} \pm6.3\mathrm{(syst.)}$~MeV/c$^2$ assuming an S-wave relativistic BW lineshape for the peak, with a statistical significance above $5\sigma$. The second structure was observed at a mass of $4313.8\pm5.3\mathrm{(stat.)}\pm 7.3\mathrm{(syst.)}$~MeV/c$^2$, however its significance could not be reliably determined because of possible reflections from two-body decays. A summary of the  masses and widths measured by CMS and other experiments is given in Table~\ref{tab:Y4140} for both structures. 
\begin{table}[htdp]
\caption{Summary of the mass and width measurements of the Y(4140) and Y(4274) structures.}
\begin{center}
\begin{tabular}{c|cc|cc}
Experiment & $\mathrm{M_1}$~(MeV/c$^2$) & $\Gamma_1$~(MeV/c$^2$) & $\mathrm{M_2}$~(MeV/c$^2$) & $\Gamma_2$~(MeV/c$^2$) \\ \hline
Belle & \multicolumn{2}{c|}{not observed} & $4350^{+4.6}_{-5.1}\pm0.7$ & $13^{+18}_{-9}\pm 4$ \\
CDF   & $4143.0^{+2.9}_{-3.0}\pm 0.6$ & $15.3^{+10.4}_{-6.1}\pm 2.5$ & $4274.4^{+8.4}_{-6.7}\pm 1.9$ & $32.3+21.9\pm 7.6$ \\
CMS   & $4148.0\pm 2.4\pm 6.3$ & $28^{+15}_{-11}\pm 19$ & $4313.8\pm 5.3\pm 7.3$ & $38^{+30}_{-16}\pm 16$ \\
D0    & $4159.0\pm 4.3\pm 6.6$ & $19.9\pm 12.6^{+1.0}_{-8.0}$ & $4328.5\pm 12.0$ & 30 (constrained) \\
LHCb  & \multicolumn{2}{c|}{not observed} & \multicolumn{2}{c}{not observed}
\end{tabular}
\end{center}
\label{tab:Y4140}
\end{table}%

The D0 collaboration analysed an integrated luminosity of 10.4~fb$^{-1}$ collected in $\mathrm{p\bar{p}}$ collisions. A binned maximum-likelihood fit to the $J/\psi\phi\mathrm{K^+}$ invariant mass distribution yielded 215 candidates. Figure~\ref{fig:CMS-D0-Y4140} ({\it right}) shows the $\mathrm{B^+}$ yield per 30~MeV corrected for detector efficiency and acceptance as function of the $J/\psi\phi$ invariant mass. A structure near threshold was observed and its significance estimated from a binned least-squared fit assuming a BW signal shape with unconstrained mass and width, convoluted with the detector resolution. The background shape is given by the non-resonant three-body phase space continuum. $52\pm 19$ signal events were obtained from the fit corresponding to a statistical significance of $3.1\sigma$, and the excess was interpreted as the Y(4140). The data also supported the presence of a second structure around 4300~MeV/c$^2$ but it was not possible to reliably fit the data with an unconstrained width. The significance of the second structure was found to be $1.7\sigma$ by varying the width between 10 and 50 MeV. The mass and widths of the first structure as measured by D0 are reported in Table~\ref{tab:Y4140} and the relative branching fraction $\mathrm{Br(B^+ \to Y(4140)K^+) / Br(B^+\to J/\psi\phi K^+)}$ extrapolated over the kinematically allowed $J/\psi\phi$ mass range was $19\pm 7\mathrm{(stat)} \pm 4\mathrm{(syst)}\%$.

In summary, since its first observation at CDF the Y(4140) and Y(4274) narrow structures in the $J/\psi\phi$ system have been studied by four different experiments at hadron and $e^+e^-$ colliders. The structure near threshold has been observed by three experiments and the statistical significance is above $5\sigma$ in two cases. The mass and width measurements agree well within the statistical and systematic uncertainties. The LHCb collaboration reports no excess near threshold, in disagreement with the CDF observation at $2.4\sigma$ level. A second structure has been observed by four experiments, with statistical significance above $3\sigma$ in three cases. There is some tension between experiments concerning the mass measurement, with values spanning between 4274 and 4350 MeV/c$^2$. Also in this case, LHCb reports no excess using the CDF  model to fit the data. In the future, Dalitz analyses of the final state particles and determination of $J^{PC}$ can be expected by exploiting the large datasets collected by the LHC experiments.

\end{document}